# Transmission des efforts dans un milieu composé de grains non sphériques

E. AZEMA(1), F. RADJAI(1), R. PEYROUX(1), G. SAUSSINE(2)

(1) LMGC, CNRS-Université Montpellier II, 34095 Montpellier cedex 05
azema@lmgc.univ-montp2.fr, radjai@lmgc.univ-montp2.fr, peyroux@lmgc.univ-montp2.fr

(2) SNCF, Direction de d'Innovation et de la Recherche, 45 rue de Londres, 75379 PARIS Cedex 08
gilles.saussine@sncf.fr

**Résumé :** Dans un milieu granulaire, les efforts de cisaillement sont entièrement transmis par l'intermédiaire d'un réseau « fort » de contacts, assisté par un ensemble de contacts appartenant à un réseau « faible ». Cette organisation non triviale des forces a toujours été établie uniquement pour un milieu granulaire composé de grains isométriques comme les disques ou sphères. L'objectif du travail proposé est de mettre en évidence, par des simulations numériques, l'influence de la forme des grains sur le comportement tant macroscopique que microscopique (la texture) d'un milieu granulaire soumis à un effort de cisaillement. Nous allons nous focaliser sur des grains de forme pentagonale simulés à l'aide de la méthode de Dynamique des Contacts. Nous avons réalisé des simulations de compression biaxiale de systèmes composés de particules pentagonales et circulaires possédant la même distribution granulométrique. La résistance au cisaillement est plus élevée pour les pentagones. Un résultat contre-intuitif de cette étude est l'observation d'une anisotropie structurale plus faible du système de pentagones par rapport aux disques. Nous montrons que la résistance au cisaillement plus élevée des pentagones est due à une anisotropie plus importante des forces qui rattrape ainsi la faiblesse de l'anisotropie structurale. Nous retrouvons, pour ce système, le même comportement en réseaux « faible » et « fort » que pour les disques. Mais les simulations montrent sans ambiguïté l'existence d'une classe de forces « très faibles » portées principalement par des contacts face-sommet. Par ailleurs, le réseau « fort » est essentiellement constitué de contacts face-face qui forment des chaînes de forces en zig-zag.

**Mots Clés.** Milieux granulaires, texture, anisotropie, chaîne de forces, effet de forme

## I) Introduction

La transmission des efforts dans un milieu granulaire a suscité un intérêt particulier depuis l'observation des inhomogénéités des forces par Dantu ([1]). La distribution des forces a été analysée grâce à des expériences et des simulations numériques par éléments discrets ([2,6]). Les forces appartiennent à deux classes distinctes qui contribuent différemment à l'anisotropie (texture), à l'effort de cisaillement, et à la dissipation ([7]). En particulier, l'effort de cisaillement est entièrement transmis par l'intermédiaire d'un réseau « fort » de contact, qui est soutenu par les contacts « faibles ». Les fonctions de distribution de probabilité (pdf) sont exponentielles pour les forces fortes, et uniformes ou en loi de puissance pour les forces faibles ([2-6]).

Mais ces études ont été essentiellement réalisées pour des grains modèles, i.e circulaires ou sphériques. Or, dans divers domaines de la science et de l'ingénierie, les grains ont des formes variées. En particulier, les grains anguleux sont très souvent utilisés comme matériau de construction nécessitant une résistance au cisaillement plus élevée. Un exemple connu est le ballast qui constitue un composant essentiel des voies ferrées ([8]). L'analyse du comportement d'un système de grains anguleux permet donc de déterminer la robustesse des comportements établis pour les grains sphériques en vue d'application aux matériaux réels.

Dans cet article, nous étudions la transmission des efforts dans un système polydisperse de particules pentagonales régulières simulé par la méthode de Dynamique des Contacts ([9]). Parmi les polygones réguliers, les pentagones ont le plus bas nombre de côtés, sans les propriétés pathologiques des triangles ou des carrés. Ce système sera notamment comparé avec un système composé de disques. Les deux systèmes sont soumis à une compression biaxiale et analysés en termes des propriétés de résistance et de texture. On présente d'abord brièvement le protocole numérique, puis on compare la résistance et la dilatance des deux systèmes (polygones et disques). Dans la section IV, on analyse l'organisation des contacts et des forces (texture). La distribution des efforts est étudiée en détail dans la section V.

## II) Procédures numériques

Dans le cadre de ce travail nous avons choisi une méthode par éléments discrets basée sur l'approche de la Dynamique des Contacts en considérant des grains convexes parfaitement rigides. Les contacts face-face entre grains anguleux sont traités en considérant deux points de contacts distincts situés aux extrémités des segments en contact. Les simulations présentées ont été menées à l'aide de la plate-forme logicielle LMGC90 ([10]), développé au Laboratoire de Mécanique et Génie Civil de Montpellier. Nous avons considéré deux systèmes ayant la même distribution granulométrique (50% de diamètre de 2,5 cm, 34 % de diamètre de 3,75 cm et 16% de diamètres de 5 cm) composés respectivement de 14 400 pentagones réguliers, noté S1, et de 10 000 disques, noté S2. On adopte pour l'ensemble des simulations une loi de contact inélastique associée à loi de frottement de Coulomb en fixant un coefficient de frottement de 0.4 entre grains et un coefficient nul pour le contact entre les grains et les parois. Les deux systèmes sont préparés suivant le même protocole. Les grains sont déposés sur un réseau géométrique, puis soumis à une compression isotrope par l'application d'une pression de $10^4$ Pa sur les parois supérieure et droite. On obtient ainsi pour les deux échantillons un rapport d'aspect $h/l \approx 2$, $l$ et $h$ étant la longueur et la hauteur de l'échantillon. La compacité initiale est de 0,80 pour S1 et de 0,82 pour S2.

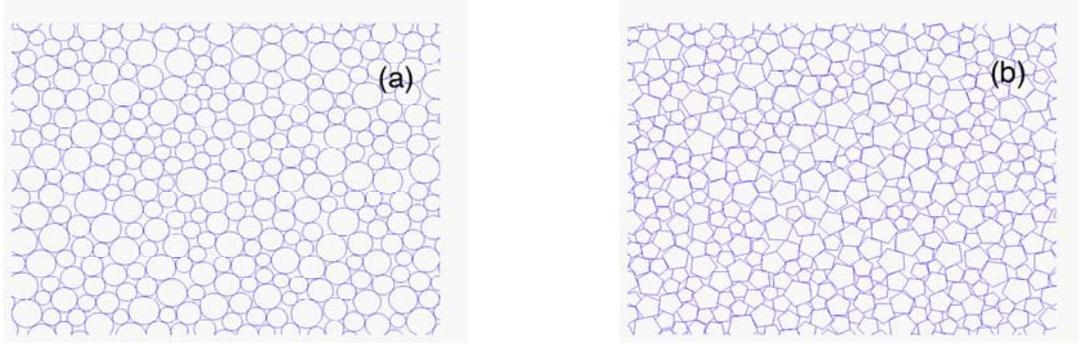

*Figure 1 : Portions des systèmes S2 (a) et S1 (b).*

Les deux systèmes sont ensuite soumis à une compression verticale en imposant une vitesse constante de 1 cm/s sur la paroi supérieure, et une pression de $10^4$ Pa sur les parois latérales. Les calculs sont menés jusqu'à obtenir une déformation verticale de 20%. Pour ce taux de compression, le taux d'énergie dissipée est négligeable par rapport à la pression de confinement, ce qui correspond à une déformation globalement quasistatique, même si localement on observe toujours des effets dynamiques, tels que des réarrangements brutaux.

### III) Comportement macroscopique

Dans cette partie, on compare les évolutions de la contrainte de cisaillement et de la dilatance des systèmes S1 et S2. Pour le calcul du tenseur de contrainte, on considère le « moment tensoriel » $M^i$ pour chaque particule $i$, définit par ([9,11]):

$$M^i{}_{\alpha\beta} = \sum_{c \in i} f_\alpha^c r_\beta^c,$$

avec $f_\alpha^c$ la composante $\alpha$ de la force exercée sur la particule $i$ au contact c, et $r_\beta^c$ la composante $\beta$ du vecteur position du même contact c. On peut montrer alors que le tenseur de contrainte d'un milieu granulaire de volume V se met sous la forme suivante ([9,11]) :

$$\sigma = \frac{1}{V} \sum_{i \in V} M^i,$$

On extrait la pression moyenne du système $p = (\sigma_1 + \sigma_2)/2$, ainsi que la contrainte déviatorique $q = (\sigma_1 - \sigma_2)/2$, avec $\sigma_1$ et $\sigma_2$ les valeurs principales. Dans nos simulations, la direction de déformation est verticale, on définit alors la déformation verticale cumulée $\varepsilon_1$ et la déformation volumique $\varepsilon_p$ par :

$$\varepsilon_1 = \int \frac{dh}{h} = \ln(1 + \frac{\Delta h}{h_0}), \quad \text{et} \quad \varepsilon_p = \int \frac{dV}{V} = \ln(1 + \frac{\Delta V}{V_0})$$

avec $h_0$ la hauteur initiale, $\Delta h = h - h_0$ la variation de hauteur, $V_0$ le volume initial et $\Delta V = V - V_0$ la variation de volume.

La figure 2 montre l'évolution de la contrainte déviatorique normalisée $q/p$ pour S1 et S2 en fonction du cisaillement $\varepsilon_q = \varepsilon_1 - \varepsilon_2$. On retrouve pour le système S2 un comportement classique caractérisé par une croissance rapide suivie d'une légère décroissance plus lente puis un plateau ([12-13]). Dans le cas des polygones, on observe un comportement similaire jusqu'à 5% de déformation, avec un pic où q/p=0.35, plus élevé que pour les disques. Au-delà on remarque que le déviateur reste autour de cette valeur pic sans diminuer. La figure 3 représente la variation volumique $\varepsilon_p$ en fonction de $\varepsilon_q$ pour S1 et S2. Les deux systèmes se dilatent et tendent vers une déformation isochorique à grandes déformations. À faible déformation, S1 se dilate moins que S2 et inversement à grandes déformations. Ceci s'explique par le fait que les particules polygonales peuvent former des pores plus importants que les particules circulaires. Le niveau élevé de $q/p$ reflète l'organisation microstructurale du système de pentagones. Ce point est étudié dans les sections suivantes.

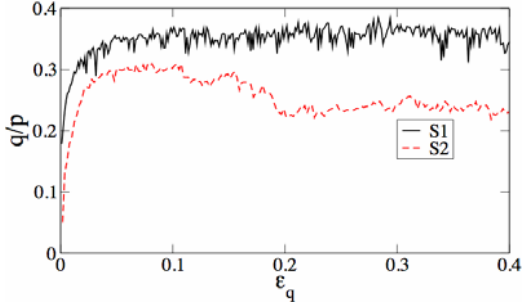
Figure 2 : Evolution de q/p en fonction de $\varepsilon_q$ pour S1 et S2.

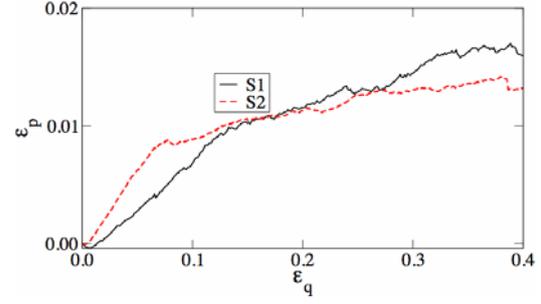
Figure 3 : Evolution de $\varepsilon_p$ en fonction de $\varepsilon_q$ pour S1 et S2.

### IV) Anisotropies des contacts et des forces

La résistance d'un milieu granulaire est en général attribuée à la formation de structures anisotropes durant la déformation, due au frottement entre grains ainsi qu'à la forme des particules ([13,15,16,17]). En deux dimensions, une manière de caractériser cette anisotropie est de considérer la fonction de probabilité $P_\theta(\theta)$, où $\theta$ représente l'orientation de la normale au contact $\underline{n}$. On définit alors le « tenseur de fabrique » par ([12]) :

$$F_{\alpha\beta} = \frac{1}{\pi} \int_0^\pi n_\alpha(\theta) n_\beta(\theta) P_\theta(\theta) d\theta \equiv \frac{1}{N_c} \sum_{c \in V} n_\alpha n_\beta,$$

où $N_c$ est le nombre total de contacts, et $\alpha$ et $\beta$ désignent les composantes du vecteur normal. L'amplitude de l'anisotropie est alors définie par $a = 2(F_1 - F_2)$, ou $F_1$ et $F_2$ sont les valeurs principales de $F$. Pour une direction $\theta'$ fixée, on peut également définir l'amplitude $a' = 2(F_1 - F_2)\cos 2(\theta_F - \theta')$, où $\theta_F$ est la direction principale de $F$. Nous avons $a' = a$ lorsque $\theta' = \theta_F$. Autrement, $a'$ permet de rendre compte de la direction de l'anisotropie.

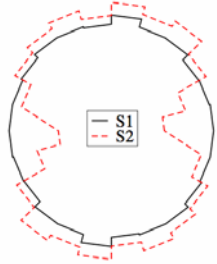
Figure 4 : Représentation polaire de $P_\theta(\theta)$ pour S1 et S2.

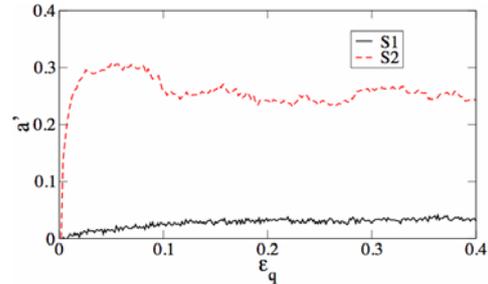
*Figure 5 : Evolution de $a'$ en fonction de $\varepsilon_q$ pour S1 et S2.*

Sur la figure 4 est tracée une représentation polaire de $P_\theta(\theta)$, pour $\varepsilon_q = 0.3$. On observe une distribution faiblement anisotrope pour S1, alors que S2 présente une forte anisotropie. La faible anisotropie de S1 est surprenante étant donnée son importante résistance au cisaillement (fig. 2). Ceci se confirme si on trace l'évolution de $a'$ durant la déformation (fig. 5). Dans les deux cas, $a'$ croît de 0 (état initial isotrope) vers une valeur seuil. La faible valeur de l'anisotropie des contacts résulte de l'organisation des contacts face-sommet et face-face dans le système S1. Pour décrire complètement la microstructure des systèmes S1 et S2, il faut s'intéresser également à l'anisotropie des forces normale $a_n$ et tangentielle $a_t$. Pour cela, définissons les « tenseurs de force » de la manière suivante ([7]) :

$$H^{(n)}_{\alpha\beta} = \int_0^\pi \langle f_n \rangle(\theta) n_\alpha n_\beta d\theta \quad \text{et} \quad H^{(t)}_{\alpha\beta} = \int_0^\pi \langle f_t \rangle(\theta) n_\alpha n_\beta d\theta,$$

où les fonctions $\langle f_n \rangle(\theta)$ et $\langle f_t \rangle(\theta)$ représentant les valeurs moyennes des forces normales et tangentielles pour la direction $\theta \in [\theta - \Delta\theta/2, \theta + \Delta\theta/2]$ des contacts, avec $\Delta\theta$ un incrément d'angle. Par définition, les fonctions $\langle f_n \rangle(\theta) = \langle f \rangle \{1 + a_n \cos 2(\theta - \theta')\}$ et $\langle f_t \rangle(\theta) = \langle f \rangle a_t \sin 2(\theta - \theta')$ sont $\pi$-périodiques, et en les approximant par leur développement au deuxième ordre, on montre que :

$$a_n = 2 \frac{H_1^{(n)} - H_2^{(n)}}{H_1^{(n)} + H_2^{(n)}} \quad \text{et} \quad a_t = 2 \frac{H_1^{(t)} - H_2^{(t)}}{H_1^{(t)} + H_2^{(t)}}$$

où $H_i^{(n)}$ et $H_i^{(t)}$ sont les valeurs propres respectivement de $H^{(n)}$ et $H^{(t)}$.

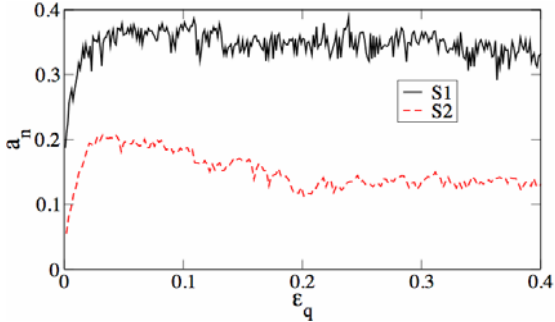
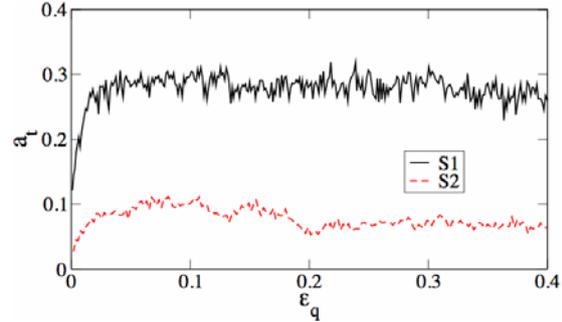

Figure 6 : Evolution de $a_n$ en fonction de $\varepsilon_q$ pour S1 et S2.  Figure 7 : Evolution de $a_t$ en fonction de $\varepsilon_q$ pour S1 et S2.

Sur les figures 6 et 7 on trace l'évolution de $a_n$ et $a_t$ en fonction de $\varepsilon_q$ pour S1 et pour S2. Contrairement à l'anisotropie des contacts (fig. 5), l'anisotropie des forces dans le cas des pentagones reste nettement supérieure à celle des disques. Ceci montre l'aptitude du système composé de pentagones à former de grandes chaînes de fortes forces. En effet, à cause de la géométrie des pentagones, c'est-à-dire en l'absence de cotés parallèles, les chaînes de forces sont la plupart du temps en forme de zig-zag (fig 8). La stabilité de telles structures impose une forte mobilité des forces tangentielles, dans ce cas nous avons bien $a_t$ très proche de $a_n$, alors que dans S2, $a_t$ est presque moitié moindre que $a_n$.

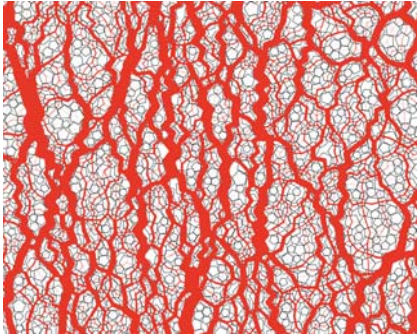
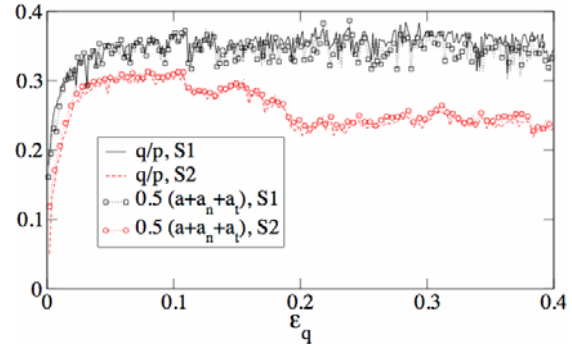

Figure 8 : Carte des forces normales pour S1. L'épaisseur du trait est proportionnelle au module de la force.  Figure 9 : Evolution de q/p en fonction de $\varepsilon_q$ pour S1 et S2.

Les anisotropies $a$, $a_n$ et $a_t$ sont des descripteurs permettant de rendre bien compte de la microstructure granulaire et de la transmission des efforts car ils peuvent être reliés à la grandeur macroscopique $q/p$. En effet, on peut montrer que ([14]) :

$$\frac{q}{p} \approx \frac{1}{2}(a + a_n + a_t)$$

où les produits croisés entre les anisotropies ont été négligés. La figure 9 montre que cette décomposition de $q/p$ est tout à fait valable pour les disques et pour les polygones. Une conséquence remarquable est de mettre en évidence l'origine de la résistance dans le cas des pentagones : pour les disques, c'est l'anisotropie des contacts qui contrôle de manière significative la stabilité du système alors que pour les polygones, la résistance élevée du matériau est assurée par l'anisotropie importante des forces tangentielles et normales.

## V) Distribution des efforts

La forte inhomogénéité des forces est un phénomène connu dans les milieux granulaires ([1-6]). Mais les études ont essentiellement porté sur des grains sphériques ou cylindriques. La probabilité de distribution des efforts est caractérisée par deux propriétés : 1) la densité de probabilité décroît de manière exponentielle pour les forces supérieures à la force moyenne, 2) pour les forces inférieures à la force moyenne, la densité de probabilité ne décroît pas vers zéro lorsque la force tend vers 0. Les études réalisées montrent que dans le cas de forces faibles, la distribution est sensible aux détails de la microstructure ([12]). L'observation commune est qu'il y a un grand nombre de contacts transmettant des forces très faibles, ce qui est une signature de l'effet de voûte. De ce point de vue, on peut penser que la forme angulaire des particules influencera principalement la distribution des forces faibles.

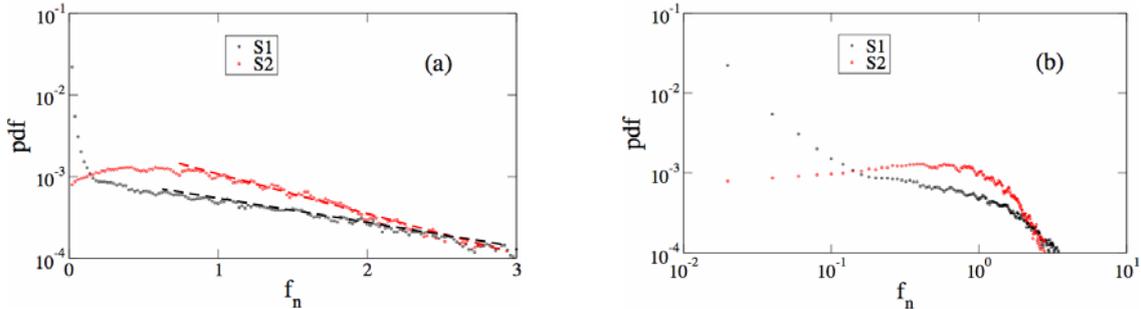

Figure 10 : Densité de probabilité des forces normales pour S1 et S2 en échelle log-linéaire (a) et en échelle log-log (b).

La distribution des forces normales moyennées par la force moyenne $\langle f_n \rangle$ est représentée en échelle log-linéaire (fig. 10a) et en échelle log-log (fig. 10b). Dans les deux systèmes, le nombre de forces fortes (supérieures à $\langle f_n \rangle$) décroît de manière exponentielle :

$$f_n \propto e^{-\alpha_1 f_n / \langle f_n \rangle} \quad pour \quad S1, \quad et \quad f_n \propto e^{-\alpha_2 f_n / \langle f_n \rangle} \quad pour \quad S2,$$

avec $\alpha_1 \approx 0,74$ et $\alpha_2 \approx 1,4$. Une faible valeur de $\alpha_1$ signifie que la distribution est plus étalée dans le cas des polygones que dans le cas des disques. Dans S2, la distribution est quasiment uniforme pour les faibles forces ($f_n < \langle f_n \rangle$). En revanche, dans le cas de S1, on observe une distribution uniforme uniquement pour $0,1\langle f_n \rangle < f_n < \langle f_n \rangle$, ce qui représente environ 30% des contacts dans le système. Le nombre de « très faibles » forces dans S1, c'est à dire pour $f_n < 0,1\langle f_n \rangle$ diminue plus rapidement qu'une loi de puissance et représente également environ 30% des contacts. La présence d'un nombre non négligeable de contacts « très faibles » dans le système formé de pentagones est une signature claire des effets de voûte ou arches comme on peut le voir sur la figure 11.

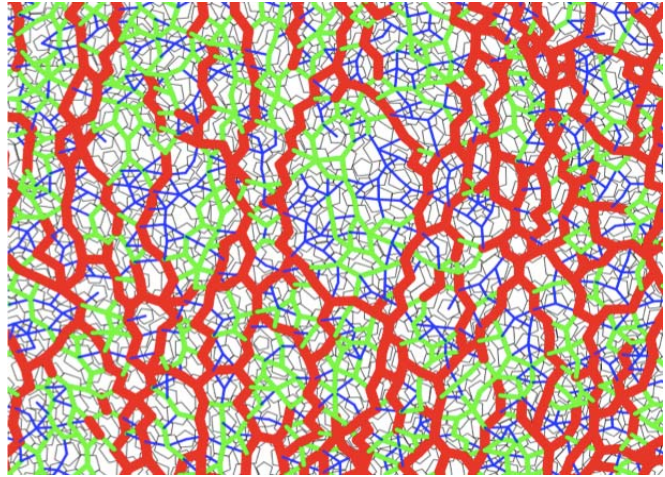

Figure 11 : Carte des contacts. En rouge les contacts forts, en verts les contacts faibles et en bleu les contacts très faibles.

## VI) Conclusions

Dans cet article, nous avons analysé, à partir des simulations numériques 2D, l'organisation microstructurale de deux milieux granulaires, soumis à une compression biaxiale : un système composé de grains circulaires et un autre de grains pentagonaux. D'un point de vue macroscopique, les pentagones permettent d'obtenir une résistance nettement supérieure à celle obtenue dans le cas de grains circulaires. Nous avons étudié l'anisotropie des contacts, l'anisotropie des forces normales et l'anisotropie des forces tangentielles et leurs liens avec la résistance au cisaillement. En particulier, nous avons montré que dans le cas des polygones, l'anisotropie des contacts est inférieure à celle des disques et que la résistance au cisaillement du système des pentagones est plutôt liée à la forte anisotropie des forces induites principalement par l'organisation des contacts face-face. Une étude détaillée a été réalisée sur la distribution des efforts dans les deux systèmes. On observe une distribution exponentielle des forces fortes dans les deux cas. En revanche, une classe de contacts très faibles apparaît distinctement, composée essentiellement de contacts face-sommet.

Cette étude montre l'importance de la forme des particules sur la transmission des forces. En particulier, dans le cas des pentagones, l'anisotropie structurale n'est pas la seule origine de la résistance au cisaillement. Les chaînes de forces impliquant des contacts face-face semblent jouer un rôle majeur dans ce système. D'autres études sont en cours afin d'analyser plus en détail la connectivité des pentagones et l'effet des contacts face-face et face-sommet. Une étude 3D est actuellement en cours pour comparer des assemblages de grains polyhédriques avec des systèmes de sphères pour une meilleure compréhension des mécanismes de transmission d'efforts dans les milieux naturels.

## VI) Références